\documentstyle{article}

\begin{document}
\centerline{RELATIVISTIC SPIN-STATISTICS CONNECTION AND 
KALUZA-KLEIN SPACE-TIME}
\vskip 2cm
\centerline{J. Anandan}
\centerline{Institute for Advanced Studies}
\centerline{Hebrew University of 
Jerusalem}
\centerline{Givat Ram, Israel 91904}
\centerline{and}
\centerline{Department of Physics and Astronomy}
\centerline{University of 
South
Carolina}
\centerline{Columbia, SC 29208, USA.}
\centerline{E-mail: jeeva@sc.edu}

\begin{abstract}
The non-relativistic formalism introduced by Berry and Robbins that
naturally incorporates the spin-statistics connection is generalized
relativistically. It is then extended to an arbitrary Kaluza-Klein
space-time by a suitable generalization of the Schwinger treatment of
angular momenta.  This leads, in this approach, to the inclusion of the
`internal' quantum numbers in the spin-statistics connection on an equal
footing with spin.
\end{abstract}
\vskip 1cm
\vskip .5cm
\noindent{May 9, 98. Revised Sept. 25, 98.}
\vskip .3cm
\noindent{Physics Letters A 248 (1998) 124-130, hep-th/9806070.}
\newpage
\par

Two particles are said to be identical or indistinguishable if in any 
state of any physical system containing these particles, interchanging 
them would not lead to a state that could be experimentally 
distinguished, even in principle, from 
the original state. Formally, we may say that both particles belong 
to the same irreducible representation of the symmetry group of 
physics. Then the two particles have the same quantum numbers 
that couple to external fields, such as mass, spin, color, weak isospin 
and hypercharge. Since the only way to observe the two particles is 
through their interactions, they are then indistinguishable.

This operational view of indistinguishability suggests, but does {\it not}
imply, that if two 
identical particles in any 
state of a system are 
interchanged then the same state, that must therefore be 
represented by the same state vector multipled by a phase factor, 
may be 
obtained. Formally, the states carry one dimensional 
representations of the permutation group. There are only two such 
representations, the symmetric and anti-symmetric, in which the 
above mentioned phase factor is $+1$ and $-1$, respectively. 

As is well known, the choice between them is made by the spin-statistics
theorem 
that states that for integer spin particles this phase factor 
is 
$+1$ and for half-integer spin particles it is $-1$. This 
has 
been derived from quantum field theory in the following two ways, 
both of which rely on the existence of anti-particle fields: 
a) The energy spectrum for the Dirac field would 
not be bounded below unless anti-commutation relations are used 
for the creation and annhilation operators in the Hamiltonian. b) 
The commutators of fields at two points that are space-
like separated vanish if the fields have integer spin and do not 
vanish if they have half-integer spin, whereas the reverse is true for 
anti-commutators. So, in order for causality to be valid we are forced 
to 
adopt commutators for integer spins and anti-commutators for half-
integer spins. 
This immediately gives the spin-statistics theorem 
\cite{we1995}.

The fact that there cannot be such a compelling theorem in non 
relativistic quantum mechanics is seen as follows: 
Indistinguishability implies that the Hamiltonian is invariant under 
interchange of any two identical particles. (The converse is not true: 
we can have a Hamiltonian for two distinguishable particles that is 
symmetric under exchange. Therefore, indistinguishablility is {\it 
not} synonymous with the Hamiltonian having this symmetry.) 
Suppose at some initial time, in violation of this theorem, the state 
of 
a set of identical half-integer (integer) spin particles is assumed to 
be a symmetric 
(anti-symmetric) wavefunction. Then, owing to the Hamiltonian 
being 
symmetric with respect to interchange of the identical particles, this 
symmetry (anti-symmetry) of the wavefunction would be preserved 
in time and there would be no inconsistency. This is also true in relativistic quantum mechanics, but in non-relativistic quantum mechanics there are no additional field theoretic
considerations which force the spin-statistics connection.

But against this must be considered several arguments which claim 
to obtain the spin-statistics connection in non-relativistic quantum 
mechanics \cite{du1998}. A particularly popular simple argument 
due to 
Feynman \cite{fe1987,fi1955}, considers 
interchanging the ends of a belt, supposed to represent two identical 
particles, without rotating them. Then the belt acquires a twist, 
which may be eliminated by rotating one end by $2\pi $ radians. 
This suggests that exchanging the particles is in some sense 
equivalent to rotating one of them by $2\pi$ radians, which results 
in a sign change for half-integer spins but no sign change for integer 
spins. 

This argument is the intuitive basis of a recent paper by Berry and 
Robbins \cite{be1997}. They have constructed an elegant and simple 
formalism which naturally yields the sign change $(-1)^{2S}$ when 
the position and spin states of two 
identical particles, with spin $S$, are exchanged, without any 
reference to relativity or quantum field theory. The fundamental 
new idea in their paper is the introduction of an exchange operator 
that interchanges the spin states without the 
above mentioned twist in the hypothetical `belt' connecting them. 
Their work, however, raises the 
following two problems: 1) The proof of the spin-statistics theorem 
in 
relativistic quantum field theory is valid but 
appears to be independent of the Berry-Robbins argument. Can there 
be two independent proofs of the same result in physics? 2)  Their 
result \cite{be1997} unextended implies the statement that the 
spatial-spin part of the 
wavefunction is always symmetric for integer spins and 
anti-symmetric for 
half integer spins. But this need not be true if the wavefunction has 
other 
degrees of freedom. (In fact the violation of the above statement, 
together 
with the spin-statistics theorem led to the discovery of color.) So, it 
is 
necessary to include the other quantum numbers in this argument.

Another reason for examining the Berry-Robbins formalism is that 
their exchange operator may change the spins of the individual 
particles while keeping the total spin the same. So, it performs an 
interesting `supersymmetry' transformation on 
the individual particles. But in order to relate this to the usual 
supersymmetry which is intimately related to relativity, it would 
appear necessary to make their formalism relativistic.

In this paper, I shall introduce the Berry-Robbins formalism into 
relativistic quantum theory. The new formalism will use the 
structure of the Lorentz group, and not just its rotation subgroup, in 
an essential way which makes it close to the quantum field theoretic 
proof, mentioned above. This suggests a link 
between the Berry-Robbins formalism and relativistic quantum field 
theory which may overcome problem (1). There is no velocity of light 
$c$ in the spin-statistics theorem. It 
obviously exists in the non-relativistic limit. So, it may well be that 
non-relativistic 
quantum mechanics `remembers' the spin-statistics connection in 
relativistic quantum field theory as I shall argue later. On the other 
hand, because this theorem has no $c$ which would 
have enabled us to take the usual non-relativistic limit through 
$c\rightarrow \infty$, it is not possible to regard the non-
relativistic 
argument as an approximation to the relativistic one. The connection 
between the two, if it exists, must be a different one!
I then extend this 
formalism to fields on an arbitrary Kaluza-Klein space-time. This 
results in the inclusion of ``internal'' variables into the argument as 
required in problem (2) above. As byproducts, the Schwinger formalism for representations 
of $SU(2)$ is generalized to the Lorentz group and all $SU(n)$.

As is well known, the generators of the Lorentz group or its covering 
group $SL(2,C)$, denoted $J_i$ and $K_i, i=1,2,3$ generating 
rotations and boosts respectively, satisfy the Lie algebra relations
$$[J_i,J_j]=i\epsilon_{ijk} J_k, [K_i,K_j]=-i\epsilon_{ijk} J_k , 
[J_i,K_j]=i\epsilon_{ijk} K_k. $$
On defining ${\bf A}={1\over 2}({\bf J}+i{\bf K})$ and ${\bf 
B}={1\over 2}({\bf J}-i{\bf K})$, equivalently
\begin{equation}
[A_i,A_j]=i\epsilon_{ijk} A_k, [B_i,B_j]=i\epsilon_{ijk} B_k , 
[A_i,B_j]=0. 
\label{lorentz}
\end{equation}
Therefore, ${\bf A}$ and ${\bf B}$ generate two commuting $SU(2)$  
groups, called the left and right handed groups, denoted here by 
$SU(2)_+$ and $SU(2)_-$. Hence, 
all the irreducible representations of $SL(2,C)$ are the same as all 
the pairs of irreducible representations of $SU(2)_+$ and $SU(2)_-$. 
So, they may be labeled by $(A,B)$, where $A$ and $B$ are the 
``spin'' quantum numbers, that are integers or half-integers, for an 
arbitrary pair of irreducible representations of $SU(2)_+$ and 
$SU(2)_-$. But in general, each irreducible representation $(A,B)$ of 
$SL(2,C)$ is reducible with respect to the physical rotation subgroup 
generated by
\begin{equation}
{\bf J}={\bf A}+{\bf B},
\label{spin}
\end{equation}
which will be denoted by $SU(2)_J$. Applying the usual laws for 
adding ``angular momenta'' for the $SU(2)_+$ and $SU(2)_-$ 
representations, the irreducible representations of $SU(2)_J$ 
contained in $(A,B)$ have spin
\begin{equation}
j=A+B, A+B-1, ...., |A-B|.
\label{reps}
\end{equation}
An invariant for all these spin representations  
is $(-1)^{2j}= (-1)^{2(A+B)}$ for all $j$ in (\ref{reps}).

I shall now generalize the Schwinger formalism 
\cite{sc1965} for non-relativistic spin to the 
representations of the 
Lorentz group.  The basic idea is to associate with each of $SU(2)_+$ 
and $SU(2)_-$ a pair of
independent oscillators whose number eigenstates determine 
basis states for all representations of this group. Let $p,q,r,s$ and 
$p^\dagger, q^\dagger,r^\dagger,s^\dagger$, respectively, be the 
annhilation and 
creation operators of the four commuting oscillators. Then {\bf A} 
and 
{\bf B} may be represented by
\[ {\bf A}={1\over 2}\left( 
 \begin{array}{cc} p^\dagger & q^\dagger \end{array}
\right)
{\bf \sigma}
\left(
\begin{array}{c} p\\q \end{array}
\right)
\]
and
\[ {\bf B}={1\over 2}\left( 
 \begin{array}{cc} r^\dagger & s^\dagger \end{array}
\right)
{\bf \sigma}
\left(
\begin{array}{c} r\\s \end{array}
\right)
\]
where ${\bf \sigma}=(\sigma^1,\sigma^2,\sigma^3)$ are the Pauli spin
matrices.
Then 
$$A_z={1\over 2}(p^\dagger p-q^\dagger q),B_z={1\over 2}(r^\dagger 
r-s^\dagger s)$$
Also, $\bf A$ and $\bf B$ satisfy (\ref{lorentz}). 

From (\ref{spin}),
\[ {\bf J}={1\over 2}\left( 
 \begin{array}{cc} p^\dagger & q^\dagger \end{array}
\right)
{\bf \sigma}
\left(
\begin{array}{c} p\\q \end{array}
\right)
+{1\over 2}\left( 
 \begin{array}{cc} r^\dagger & s^\dagger \end{array}
\right)
{\bf \sigma}
\left(
\begin{array}{c} r\\s \end{array}
\right)
\]
Comaparing this with Schwinger's representation of non-relativistic 
spin \cite{sc1965} by means of two oscillators, there has been a 
doubling of the number of degrees of freedom in the present 
relativistic case where there are four oscillators. This corresponds to 
the existence of anti-particles. It is 
interesting 
that this result has been obtained here from the structure of the 
Lorentz group, in particular from the fact that its dimension is twice 
that of the rotation group. This is very different from how anti-
particles were historically discovered, namely the existence of 
negative energy 
solutions of relativistic wave equations, which in turn is due to the 
quadratic dispersion relation $E^2 = {\bf p}^2 + m^2$.

A basis of states for the $(A,B)$ representation of $SL(2,C)$ are 
eigenstates of $A^2,A_z,B^2,B_z$. These are states with definite 
numbers of quanta: 
$|n_p,n_q,n_r,n_s>={p^\dagger}^{n_p}{q^\dagger}^{n_q}
{r^\dagger}^{n_r}{s^\dagger}^{n_s}|0>$.
Then
\begin{equation}
A={1\over 2}(n_p+n_q), B={1\over 2}(n_r+n_s) .
\label{AB}
\end{equation}

Consider now two identical particles $1$ and $2$. Their states are 
spanned by
\begin{equation}
|n_{1p},n_{2p},n_{1q},n_{2q},n_{1r},n_{2r},n_{1s},n_{2s}>=
{p_1^\dagger}^{n_{1p}}{p_2^\dagger}^{n_{2p}}
{q_1^\dagger}^{n_{1q}}{q_2^\dagger}^{n_{2q}}
{r_1^\dagger}^{n_{1r}}{r_2^\dagger}^{n_{2r}}
{s_1^\dagger}^{n_{1s}}{s_2^\dagger}^{n_{2s}}|0>.
\label{quanta}
\end{equation}
Here and later the subscripts $1$ and $2$ refer 
to the particles $1$ and $2$. Analogous to the 
Berry-Robbins exchange angular momentum define
\[ {\bf E}_p={1\over 2}\left( 
 \begin{array}{cc} p_1^\dagger & p_2^\dagger \end{array}
\right)
{\bf \sigma}
\left(
\begin{array}{c} p_1\\p_2 \end{array}
\right)
\]
Similar definitions are given for ${\bf 
E}_q,{\bf E}_r$ and ${\bf E}_s$. Define
$${\bf E}={\bf E}_p+{\bf E}_q+{\bf E}_r+{\bf E}_s .$$
Then,
$$[E_i,E_j]=i\epsilon_{ijk}E_k , [E_i,S_j]=0$$
where ${\bf S}={\bf J}_1+{\bf J}_2={\bf A}_1+{\bf A}_2+{\bf 
B}_1+{\bf B}_2$ is the total spin.

Analogous to the construction of Berry-Robbins, a new wavefunction 
of the relative coordinates $\bf r$ of the two particles may be 
defined as
\begin{equation}
{\tilde\psi}({\bf r})=U({\bf r})\psi({\bf r}),
\label{BR}
\end{equation}
where $\psi({\bf r})$ is the usual wavefunction and 
$$U({\bf r}) = \exp\{-i\theta {\bf n}({\bf r})\cdot {\bf E}\}$$
with $\theta$ being the angle between  $\bf r$ 
and the $z-$ axis, and ${\bf n}({\bf r})$ is a unit vector 
perpendicular to the $z-$axis and varying smoothly with $\bf r$. One 
such smooth choice, used by Berry-Robbins, is to require ${\bf 
n}({\bf r})$ to be perpendicular
to $\bf r$, as well. The dependence of $U({\bf r})$ on the choice of 
$z-$ axis and the direction of ${\bf n}({\bf r})$ gives a gauge 
freedom in the choice of $U({\bf r})$. Each pair of indistinguishable 
configurations 
corresponding to $\bf r$ and -$\bf r$ are identified. Requiring single 
valuedness of $\tilde\psi$ in the quotient configuration space 
then amounts to the  condition
\begin{equation}
\psi(-{\bf r})=\exp\{-i\pi {\bf n}({\bf r})\cdot {\bf E}\}\psi({\bf r})
\label{sv}
\end{equation}
in terms of the usual wave function. 

Now $\psi({\bf r})$ may be 
expanded in terms of (\ref{quanta}). Using  the commutation relations 
of the annhilation and 
creation operators,
$$\exp\{-i\pi {\bf n}\cdot {\bf E}\}{p_1}^\dagger \exp\{i\pi {\bf 
n}\cdot {\bf E}\}=e^{i\phi}{p_2}^\dagger ,
\exp\{-i\pi {\bf n}\cdot {\bf E}\}{p_2}^\dagger 
\exp\{i\pi {\bf n}\cdot {\bf E}\}=-e^{-i\phi}{p_1}^\dagger$$
where $\bf n$ is perpendicular to the $3-$axis and $\phi$ is an 
inconsequential angle that $\bf n$ makes with the $2-$axis, with 
similar relations for the $q,r$ and $s$ operators. Then, using 
(\ref{AB}) and (\ref{reps}), 
$$
\exp\{-i\pi {\bf n}\cdot {\bf 
E}\}|n_{1p},n_{2p},n_{1q},n_{2q},n_{1r},n_{2r},n_{1s},n_{2s}>
=(-1)^{2j}
|n_{2p},n_{1p},n_{2q},n_{1q},n_{2r},n_{1r},n_{2s},n_{1s}>
$$
Hence, the RHS of (\ref{sv}) is 
$(-1)^{2j}\bar\psi (\bf r)$,
where the bar denotes the exchange of spin quantum numbers of the 
two particles. This 
is the sought after spin-statistics connection.

The above argument suggests that the physical argument of 
Feynman \cite{fe1987}, mentioned above, should 
be made in the abstract three dimensional spaces on which $SU(2)_+$ 
and  $SU(2)_-$ act, which makes use of the full structure of the 
Lorentz group, and not in the physical space in which  $SU(2)_J$ acts. 
Indeed, the usual argument which gives the spin-statistics 
theorem in quantum field theory \cite{we1995} makes use of the 
transformation of the field under both $SU(2)_+$ and  $SU(2)_-$. As 
mentioned above, this makes use of the degrees of freedom of the 
anti-particle as well  the particle. Since $SU(2)_J$ is a subgroup of 
$SU(2)_+\times SU(2)_-$, the `belt' argument may also be used on 
the 
physical space on which $SU(2)_J$ acts. Hence, non-relativistic 
physics `remembers' an argument in which relativity was essentially 
involved. This may well be the long sought `missing link' between 
the 
`belt' argument, which appeared to have nothing to do with 
relativity, 
and the usual argument from relativistic quantum field theory.

The Schwinger formalism will now be extended to 
an arbitrary 
Lie group $G$, with the Lie algebra relations $[T^i,T^j]=i \sum_k {C^{ij}}_kT^k$. 
Let 
\begin{equation}
\hat T^i = \sum_{p,q=1}^\nu {a^p}^\dagger T^i_{pq}a^q ,
\label{generator}
\end{equation}
where $T^i_{pq}, p,q=1,2,...\nu$ are the matrix elements of $T^i$ in 
the fundamental representation and the annhilation and creation 
operators satisfy
\begin{equation}
[a^p,{a^q}^\dagger]=\delta^{pq}, 
[a^p,a^q]=0,[{a^p}^\dagger,{a^q}^\dagger]=0,p,q=1,2,...\nu.
\label{ho}
\end{equation}
It can then be proved, using (\ref{ho}),
$$
[\hat T^i,\hat T^j]=i\sum_k {C^{ij}}_k\hat T^k .
$$
Also, since the commuting generators $T^i_C, i=1,2,...r$ of the Cartan 
subalgebra can be made diagonal, where $r$ is the rank of $G$, write 
$T^i_{Cpq}= \lambda^i_p \delta_{pq}$. 
Therefore,
\begin{equation}
\hat T^i_C=\sum_{p=1}^\nu \lambda^i_p {a^p}^\dagger a^p , i=1,2,...r.
\label{cartan}
\end{equation}

The states,
\begin{equation}
|n^1,n^2,...n^\nu>=({{a^1}^\dagger})^{n^1}({{a^2}^\dagger})^{n^2}......
({{a^\nu}^\dagger})^{n^\nu}|0>,
\label{states}
\end{equation}
are simultaneous eigenstates of the Cartan subalgebra.
In particular, the single quantum states are
$|\chi_p>\equiv {a^p}^\dagger |0>, p=1,2,...\nu$. From 
(\ref{generator}) and (\ref{ho}), the 
matrix elements 
$<\chi_p|\hat T^i|\chi_q>=T^i_{pq}$.
Hence, the fundamental representation of $G$ acts on the vector 
space spanned by $\{ |\chi_p>\}$. Since the creation operators 
commute, the tensors (\ref{states}) may 
be obtained by taking symmetrized tensor products of the vectors 
$\{|\chi_p>\}$. 

All irreducible representations of $G$ may be obtained by 
constructing the vector space $V$ spanned by tensor products of  the 
symmetric tensors (\ref{states}) and reducing the representation of $G$ that 
acts on $V$. Shmuel Elitzur has suggested adding another index, say 
$\tau$, to the creation and annhilation operators to represent the 
position of each symmetric tensor in the last mentioned tensor product on which 
they act. These symmetric tensors then correspond to the rows of a Young 
tableau (see, for example, \cite{we1931}). I.e. each value of $\tau$ 
corresponds to a particular row of the Young tableau. By anti-
symmetrizing the columns of the Young tableau, irreducible 
representations of the permutation group acting on the vector space 
spanned by the tensors of this Young tableau are obtained. I shall
therefore 
make the creation (annhilation) operators with different values of 
$\tau$ anti-commute. The Young tableau has at most $\nu$ rows 
because anti-symmetrizing more than $\nu$ elements gives zero. 
Therefore, $\tau$ takes values $1,2,...\nu$. Hence, (\ref{generator}) 
and (\ref{cartan}) may be generalized to
\begin{equation}
\hat T^i = \sum_{\tau=1}^{\nu}\sum_{p,q=1}^\nu {a^{\tau p}}^\dagger T^i_{pq}a^{\tau q} ,
\hat T^i_C=\sum_{\tau=1}^{\nu}\sum_{p=1}^\nu \lambda^i_p 
{a^{\tau p}}^\dagger a^{\tau p} .
\label{Generators}
\end{equation}
The representation corresponding to a given Young tableaux acts on a 
vector space spanned by
$$|n^{\tau p}>\equiv |n^{11}...n^{1\nu},...n^{21}...n^{2\nu},...  
n^{\nu 1}...n^{\nu \nu} >\equiv \{\prod_{\tau=1}^\nu \prod_{p=1}^\nu 
(a^{\tau p \dagger})^{n^{\tau p}}\}|0>$$
The $\tau -$th row of this Young tableaux has $n^\tau 
=\sum_{p=1}^{\nu}n^{\tau p}$ elements, where $n^{\tau p}$ are non 
negative integers.

But this representation, corresponding to a given Young tableaux, may 
be reducible under the action of $G$. The irreducible representations 
may be extracted by contracting the tensors on which this 
representation acts with tensors that are invariant under $G$ to 
obtain invariant lower dimensional representations, and by taking 
suitable linear combination of these tensors multiplied by Kronecker 
deltas where appropriate to obtain irreducible tensors. 
Anti-symmetrization of a column with $\nu$ elements corresponds to 
contracting these indices with the epsilon tensor. If the determinant 
of the transformation matrix is $1$, e.g. for $G=SU(n)$ or $SO(d)$, this contraction is invariant and a 
lower rank tensor is obtained. Hence for special groups, the Young 
tableaux may have at most $\nu-1$ rows, or equivalently $\tau 
=1,2,...\nu-1$ only.  

For the special case of $G=SU(2)$, $\nu=2$ and therefore the $\tau$ 
index takes only one possible value and may be omitted. This 
corresponds to the Schwinger formalism for angular momentum, used above.  More 
generally for 
$G=SU(n)$, $\nu=n$ and $\tau$ takes $n-1$ possible values. Then, as is well known, 
each Young tableaux uniquely corresponds to an irreducible 
representation of $SU(n)$, because the only invariant tensor is the 
epsilon tensor which has already been used to reduce the rows of the 
Young tableaux to at most $n-1$ rows. The above treatment,  with the sum 
over $\tau$ restricted to $1,2,...n-1$, then generalizes the 
Schwinger oscillator formalism for $SU(2)$ to representations of $SU(n)$. This has the 
advantage that we would always use $n(n-1)$ oscillators, with creation operators $a^{\tau 
p \dagger}$, irrespective of the dimension of the representation.

Consider now an arbitrary Kaluza-Klein space-time \cite{ap1987} of 
dimension 
$d=2m$ or $2m+1$, where $m$ is an integer $\geq 2$. The spinor 
and 
tensor fields on this space-time transform under representations of 
the covering group of $SO(1,d-1)$. They can be built from the 
fundamental spinor representation of this group, which may be 
constructed as 
follows: First construct the $\gamma$ matrices to act on tensor 
products of $m$ spin-$1\over 2$ representations. A particular set 
that is 
anti-commuting and appropriately normalized is
$$\gamma^0 =-I\otimes \sigma^1\otimes ...\otimes I ,~~\gamma^j =i 
\sigma^j\otimes \sigma^2\otimes I\otimes...\otimes I, 
j=1,2,3$$
$$\gamma^5 =i I\otimes \sigma^3\otimes\sigma^1\otimes 
I\otimes...\otimes 
I ,~~\gamma^6 =iI\otimes \sigma^3\otimes\sigma^2\otimes 
I\otimes...\otimes I$$
$$\gamma^7 =iI\otimes \sigma^3\otimes 
\sigma^3\otimes\sigma^1\otimes I\otimes...\otimes I, ~\gamma^8 
=iI\otimes \sigma^3\otimes 
\sigma^3\otimes\sigma^2\otimes I\otimes...\otimes I$$
$$.....................................$$
There are $m$ factors in the definition of each $\gamma^M$ which 
make the 
vector space they act on $2^m$ dimensional, for either value of $d$.

The generators of this fundamental spinor representation are then
\begin{equation}
S^{MN}={i\over 4}[\gamma^M,\gamma^N], M,N=1,2,....d
\label{generators}
\end{equation}
The generators of Lorentz boosts of ordinary space-time are
$$S^{0j}={i \over 2}\sigma^j\otimes \sigma^3\otimes ....\otimes I, 
j=1,2,3
$$
while the generators of spatial rotations, i.e.  $SU(2)_J$, are
$$S^{ij}={1 \over 2}\epsilon_{ijk} \sigma^k\otimes I\otimes 
....\otimes 
I,i,j=1,2,3.$$
It is clear that rotation by 
$2\pi$ radians that is in $SU(2)_J$ gives a sign change to all vectors 
in this fundamental spinor 
representation. Other representations may be obtained by making 
tensor products of the fundamental spinor representations $n$ times, 
where $n$ is any positive integer, and 
taking the 
irreducible representations. Clearly, if $n$ is odd 
this tensor product would give only half-
integer spin fields, because 
all the elements in this tensor product undergo sign change under 
the 
above $2\pi$ rotation. Similarly, if $n$ is even then there will be no 
sign change so that only integer spin fields are obtained. Hence,
\begin{equation}
(-1)^n=(-1)^{2S},
\label{signchange}
\end{equation}
where $S$ is the spin of 
any of the irreducible representations obtained for a given $n$, 
although obviously for $n> 1, n\ne 2S$, in general.

Alternatively, the group $G$ in the above treatment of its 
representations by means of oscillators may be taken to be the 
covering group of $SO(1,d-1)$. Then $\nu =2^m$ and the last mentioned 
tensor products correspond to states with $n$ quanta. Consider now 
two identical particles. Since they are identical, consider their 
states that belong to the same Young tableau  for both particles, 
which are of the form
\begin{equation}
|n_1^{\tau p},n_2^{\tau p}>\equiv 
\{\prod_{\tau=1}^{2^m-1}\prod_{p=1}^{2^m} (a_1^{\tau p 
\dagger})^{n_1^{\tau p}}
(a_2^{\tau p 
\dagger})^{n_2^{\tau p}}\}|0>
\label{2states}
\end{equation}
where 
\begin{equation}
n=\sum_{\tau=1}^{2^m-1}\sum_{p=1}^{2^m} n_1^{\tau p} 
=\sum_{\tau=1}^{2^m-1}\sum_{p=1}^{2^m} n_2^{\tau p} ,
\label{n}
\end{equation}
where $n_1^{\tau p}$ and $n_2^{\tau p}$ are non negative integers.
The generalization of the Berry-
Robbins exchange angular momentum is 
${\bf E}=\sum_{\tau=1}^{2^m-1}\sum_{p=1}^{2^m} {\bf E}^{\tau p}$, where

\[ {\bf E}^{\tau p}={1\over 2}\left( 
 \begin{array}{cc} {a_1^{\tau p}}^\dagger & {a_2^{\tau p}}^\dagger 
\end{array}
\right)
{\bf \sigma}
\left(
\begin{array}{c} a_1^{\tau p}\\a_2^{\tau p} \end{array}
\right)
\]
which now generates exchange of Kaluza-Klein spin states.
Since each ${\bf E}^{\tau p}$ is quadratic in the creation and annhilation operators, which satisfy (\ref{ho}) for a given $\tau$ and anti-commute for different $\tau$s,  the ${\bf E}^{\tau p}$s must commute among themselves.
Using these commutation relations,
$$\exp\{-i\pi {\bf n}\cdot {\bf E}\}{a_1^{\tau p}}^\dagger \exp\{i\pi 
{\bf 
n}\cdot {\bf E}\}=\exp\{-i\pi {\bf n}\cdot {\bf E}^{\tau p}\}{a_1^{\tau p}}^\dagger \exp\{i\pi 
{\bf 
n}\cdot {\bf E}^{\tau p}\}
=e^{i\phi}{a_2^{\tau p}}^\dagger ,$$
$$\exp\{-i\pi {\bf n}\cdot {\bf E}\}{a_2^{\tau p}}^\dagger 
\exp\{i\pi {\bf n}\cdot {\bf E}\}=\exp\{-i\pi {\bf n}\cdot {\bf E}^{\tau p}\}{a_2^{\tau p}}^\dagger \exp\{i\pi 
{\bf 
n}\cdot {\bf E}^{\tau p}\}=-e^{-i\phi}{a_1^{\tau p}}^\dagger$$
where $\bf n$ is perpendicular to the $3-$axis and $\phi$ is the 
angle that $\bf n$ makes with the $2-$axis. 
It follows from (\ref{2states}), (\ref{n}) and (\ref{signchange}),
\begin{equation}
\exp\{-i\pi {\bf n}\cdot {\bf E}\}|n_1^{\tau p},n_2^{\tau p}>=
(-1)^{2S}
|n_1^{\tau p},n_2^{\tau p}> .
\label{sign}
\end{equation}

A relative coordinate wave function analogous to (\ref{BR}) may now 
be constructed on Kaluza-Klein space-time, and its single 
valuedness requires (\ref{sv}) for the usual wave function, with  the 
last defined $\bf E$, and
$\bf r$ is now the $d-1$ dimensional relative coordinate vector in 
the Kaluza-Klein space. By expanding this wave function in terms of 
(\ref{2states}) and using (\ref{sign}), the RHS of (\ref{sv}) is 
$(-1)^{2S}\bar\psi({\bf r})$. I.e.
interchanging Kaluza-Klein spins and  positions gives the 
factor $(-1)^{2S}$ to the state of two identical particles each of 
which has spin $S$.
This gives the 
{\it spin-
statistics connection in Kaluza-Klein space-time}. 

When the Kaluza-Klein space-
time is compactified by curling up the `internal' dimensions 
\cite{we1983}, this gives the spin-statistics connection in the usual 
4 dimensional space-time, as will be shown now. The wave function in the fundamental 
representation on which (\ref{generators}) 
act may be written
\begin{equation}
\Psi^{\alpha^1\alpha^2....\alpha^m}(X)=\sum_{Rf}
\psi^{\alpha^1\alpha^2}_{Rf}(x)\phi^{\alpha^3....\alpha^m}_{Rf}(y).
\label{basis}
\end{equation}
Here the Kaluza-Klein coordinates $X=(x,y)$ are split into the usual 
four 
dimensional space-time coordinates $x$ and the coordinates $y$ of 
the internal space $B$. The isometry group $H$ on $B$ is the gauge 
group in the usual space-time. $R$ labels each irreducible 
representation of $H$ which acts on the basis vectors 
$\phi^{\alpha^3....\alpha^m}_{Rf}(y)$, 
and $f$ labels a vector in this representation. Each $\alpha^i$ index 
takes two possible values. The Lorentz group acts on the pair of 
indices $\alpha^1,\alpha^2$ only, while the spinor group on $B$ that 
is induced by $H$ acts on the indices $\alpha^3,....\alpha^m$.
The usual fields on space-time are $\psi^{\alpha^1\alpha^2}_{Rf}(x)$, 
where  $(\alpha^1,\alpha^2)$ is the 4 dimensional Dirac spinor index, 
while $(R,f)$  is interpreted as the `internal' index representing 
weak isospin, hypercharge and color states, etc. 

Consider now a state of two identical particles in this fundamental 
spinor representation. It follows, from the above result in italics, 
that 
their wave function must satisfy
\begin{equation}
\Psi^{\alpha_1^1....\alpha_1^m,\alpha_2^1....\alpha_2^m}(X_1,X_2)=
-\Psi^{\alpha_2^1....\alpha_2^m,\alpha_1^1....\alpha_1^m}(X_2,X_1),
\label{antisymmetry}
\end{equation}
This wave function may be expanded in terms of the basis states 
$\phi_{Rf}$ in (\ref{basis}) as
\begin{equation}
\Psi^{\alpha_1^1....\alpha_1^m,\alpha_2^1....\alpha_2^m}(X_1,X_2)=
\sum_{R_1 f_1,R_2 f_2}
\psi^{\alpha_1^1 \alpha_1^2,\alpha_2^1 \alpha_2^2}_{R_1 f_1,R_2 
f_2}(x_1,x_2)
\phi^{\alpha_1^3....\alpha_1^m}_{R_1f_1}(y_1)
\phi^{\alpha_2^3....\alpha_2^m}_{R_2f_2}(y_2).
\label{expansion}
\end{equation}
From (\ref{antisymmetry}) and (\ref{expansion}), the usual space-
time wave function satisfies
\begin{equation}
\psi^{\alpha_1^1\alpha_1^2,\alpha_2^1\alpha_2^2}_{R_1 f_1,R_2 f
_2}(x_1,x_2)=-
\psi^{\alpha_2^1\alpha_2^2,\alpha_1^1\alpha_1^2}_{R_2 f_2,R_1 f_1}
(x_2,x_1).
\label{4}
\end{equation}

The proof for the higher dimensional representations is similar to the 
above except that in (\ref{antisymmetry})-(\ref{4}) there are now 
$2mn~~\alpha$-indices, instead of the $2m~~\alpha$-indices as in 
the above special case of $n=1$, and the $-$ sign in the RHS of 
(\ref{antisymmetry}) and (\ref{4}) is replaced by the factor 
$(-1)^n=(-1)^{2S}$ . It is emphasized that in the above treatment no 
{\it a priori} preference is given to spin over other variables 
such as weak isospin, hypercharge and color, unlike in ref. 
\cite{be1997}.

The extension of this result to more than two identical particles is 
straightforward: If the state is anti-symmetric (symmetric) with 
respect to interchange of any pair of identical particles then it must 
be totally anti-symmetric (symmetric). 

I thank Michael Berry for explaining to me the results in reference
\cite{be1997} and for useful discussions. I also thank Schmuel Elitzur and Francois Englert 
for useful 
discussions. This work was 
partially supported by NSF grant PHY-9601280.

\end{document}